\documentclass[preprint,12pt]{elsarticle}
\usepackage[cp1251]{inputenc}
\usepackage{natbib}
\usepackage{amsmath}
\usepackage{amssymb}
\usepackage{textcomp}
\usepackage{graphicx}

\DeclareMathOperator{\artanh}{artanh}

\journal{Physics Letters A}

\begin{document}

\begin{frontmatter}



\title{Charge-separated electron-hole liquid\\
in transition metal dichalcogenide heterostructures}


\author{Pavel V. Ratnikov\footnote{Email address: ratnikov@lpi.ru}}

\address{A.M. Prokhorov General Physics Institute, Russian Academy of Sciences\\ Vavilova Street 38, 119991 Moscow, Russia}

\begin{abstract}
The possibility of the formation of an electron-hole liquid in the type II heterostructures based on monolayers (bilayers) of transition metal dichalcogenides is considered. It is indicated that additional valleys in the conduction band, which are present in bilayers, are required for its observation. The binding energy of the interlayer exciton is found by the variation method. An analytical formula is found for the shape of the recombination line of the electron-hole liquid.
\end{abstract}


\end{frontmatter}


\section{Introduction}
For over ten years now, vertical (layered) heterostructures composed of various two-dimensional (2D) materials, including monolayers (1L) and bilayers (2L) of transition metal dichalcoshenides (TMDs), have been actively studied. The layers are connected by van der Waals (vdW) forces. So, such heterostructures are now commonly referred to as vdW heterostructures \cite{Geim2013}.

1L-TMD films can be considered as ideal systems for the investigation of a high-temperature electron-hole liquid (EHL). 2D nature of electrons and holes ensures their stronger interaction as compared to that in bulk semiconductors. The screening of the Coulomb interaction in monolayer heterostructures is significantly reduced, since it is determined by the permittivities of the environment. The high-temperature strongly coupled EHL has been already observed in suspended MoS$_2$ monolayers \cite{Yu2019} and ultrathin MoTe$_2$ photocells \cite{Arp2019}.

Calculations using the density functional theory \cite{Terrones2014} have shown that bilayers of different stackings have different band structures, in which the difference between direct and indirect band gaps is small, but the indirect gaps are always smaller. Thus, bilayers composed of two monolayers of the same material are indirect-gap semiconductors. Contrary, TMD heterobilayers such as 1L-WSe$_2$/1L-WS$_2$, 1L-MoSe$_2$/1L-MoS$_2$, 1L-MoSe$_2$/1L-WS$_2$ and 1L-WSe$_2$/1L-MoS$_2$ are direct-gap ones, and the band gap is approximately half that of the band gap in a monolayer \cite{Terrones2013}.

Bilayers of one material have additional valleys. For example, there are 6 electron valleys in 2L-WS$_2$ at the $\Lambda$ points (the $\Lambda$ point is the middle of the $\Gamma-K$ segment), and the spin degeneracy in them has been removed \cite{Chernikov2015}. Consequently, with respect to the properties of EHL, such a system behaves in the same way as the monolayer heterostructures considered by us, but with a different number of valleys. We have $(\nu_e,\,\nu_h)\,\rightarrow\,(3,\,1)$ for 2L-WS$_2$ (in our previous work \cite{Pekh2021}, we agreed to inscribe the multiplicity of degeneracy in spin $g_s=2$ into the formulas, and in the absence of spin degeneracy, use $\nu_e/2$ for electrons and/or $\nu_h/2$ for holes). Therefore, bilayers of one TMD do not deserve separate consideration.

According to first-principles calculations \cite{Debbichi2014}, taking into account the spin-orbit coupling and vdW interaction, a number of heterobilayers have a type II contact, when the bottom of the conduction band of the second semiconductor lies in energy below the bottom of the conduction band of the first semiconductor and the edges of the valence bands are related in the same way. Then, upon photoexcitation, the first monolayer will contain holes, and the second will contain electrons. For example, this is implemented in 1L-WS$_2$/1L-MoS$_2$ \cite{Bernardi2013}, 1L-WSe$_2$/1L-MoSe$_2$ \cite{Wang2019}, and 1L-WSe$_2$/1L-MoS$_2$ \cite{Zimmermann2021}.

In our opinion, the study of EHL in vdW heterostructures based on TMDs with spatial separation of electrons and holes is of interest not only from the theoretical side, but also from the practical one. The results can be applied to photovoltaics. In particular, a broad EHL line in the photoluminescence spectrum promotes greater light absorption and, as a consequence, greater efficiency of solar cells.

Photoexcited electrons and holes in type II TMD heterobilayers separate over monolayers on a scale of tens of femtoseconds \cite{Hong2014, Zhu2017} and form oppositely charged layers. At sufficiently low charge carrier densities, interlayer excitons are formed \cite{Rivera2015, Kim2017, Rivera2018, Kunstmann2018}. A dielectric exciton gas is transformed into a conductive charge-separated plasma when the charge carrier density exceeds the metal--insulator transition density $n_{dm}$. The metal--insulator transition has previously been observed in optically excited 1L- and 2L-WS$_2$ \cite{Chernikov2015}.

The advantage of TMD heterobilayers with spatially separated electrons and holes is that charge carriers have longer lifetimes. The intralayer exciton photoluminescence is completely suppressed in 1L-WSe$_2$/1L-MoSe$_2$ at charge carrier densities $n\leq10^{13}$ cm$^{-2}$, while the radiation from the recombination of interlayer excitons dominates \cite{Wang2019}. The peak of interlayer emission broadens significantly when the threshold $n_{dm}=3\cdot10^{12}$ cm$^{-2}$ is overcome, which proves the absence of excitons above $n_{dm}$. The reappearance of intralayer emission is also observed. Its intensity increases for densities $n>10^{13}$ cm$^{-2}$. The authors of work \cite{Wang2019} qualitatively explained this effect by the influence of a strong electric field of separated charges, sufficient to compensate for the initial difference between the band edges of MoSe$_2$ and WSe$_2$ $\sim300$ meV. However, the EHL line was not observed up to densities $\sim10^{14}$ cm$^{-2}$.

EHL is not energetically favorable in the case of charge carriers separation over monolayers for $\nu_e=2$ and $\nu_h=1$ (the number of hole valleys is taken into account as divided by 2 due to the removal of spin degeneracy in the valance band). To detect charge-separated EHL, it is necessary to increase the number of valleys. As mentioned above, the number of electron valleys is greater in 2L-TMD films than in the monolayer owing to valleys at the $\Lambda$ points. Monolayer/bilayer heterostructures are required to observe the charge-separated EHL (electrons are in the bilayer, as, for example, in the 1L-WS$_2$/2L-MoS$_2$ heterostructure). The provided below calculations confirm this.

\section{Ground state energy of EHL}
We consider a model 2D semiconductor with $\nu_e$ electron and $\nu_h$ hole valleys. We assume that both electrons and holes are degenerate in spin (we enter $g_s=2$ in the formulas, and the number of valleys with removed spin degeneracy is taken into account as divided by 2). The density of produced electron-hole ($e$-$h$) pairs is equal to $n$.

We work in a system of units with $\hbar=1$. We measure the momenta (wave vectors) $q$ in units of the Fermi momentum of electrons $q_{Fe}$. More often, we have $q_{Fe}<q_{Fh}$ (the Fermi momentum of holes), $q_{Fe,h}=\sqrt{2\pi n/\nu_{e,h}}$, since $\nu_e\geq\nu_h$ and spin degeneracy is lifted in the hole valleys at the $K$ points (we take into account the upper spin branches of the valence band).

The frequency is measured in units of $q^2_{Fe}/2m$, where $m=m_{e2}m_{h1}/(m_{e2}+m_{h1})$ is the reduced mass of an electron $m_{e2}$ in the second and a hole $m_{h1}$ in the first materials. The density of $e$-$h$ pairs and the energy are respectively measured in units of $a^{-2}_x$ and $E_x$, where $a_x=1/2m\widetilde{e}^2$ is the Bohr radius of 2D exciton and $E_x=2m\widetilde{e}^4$ is the 2D exciton binding energy, $\widetilde{e}^2=e^2/\varepsilon_\text{eff}$ and $\varepsilon_\text{eff}=(\varepsilon_1+\varepsilon_2)/2$ is the effective permittivity, $\varepsilon_1$ and $\varepsilon_2$ are low-frequency permittivities of the surrounding media \cite{Ginzburg1973}.

We also introduce the dimensionless distance between particles $r_s=\sqrt{\nu_e/\pi n}$. So we have $q_{Fe}=\sqrt{2}/r_s$.

In the case of separation of electrons and holes over the layers of the heterostructure, an electrostatic contribution is added to the energy of the $e$-$h$ system \cite{Andryushin1981}.

The ground state energy of 2D EHL per one $e$-$h$ pair is
\begin{equation}\label{E_gs}
E_\text{gs}=E_\text{kin}+E_\text{elst}+E_\text{dd}+E_\text{exch}+E_\text{corr}.
\end{equation}

The first term is the kinetic energy ($\varkappa=\nu_e/\nu_h$ and $\sigma=m_{e2}/m_{h1}$) \cite{Pekh2021}
\begin{equation}\label{E_kin}
E_\text{kin}=\frac{1+\varkappa\sigma}{1+\sigma}r^{-2}_s.
\end{equation}

\begin{figure}[t!]
\begin{center}
\includegraphics[width=12cm]{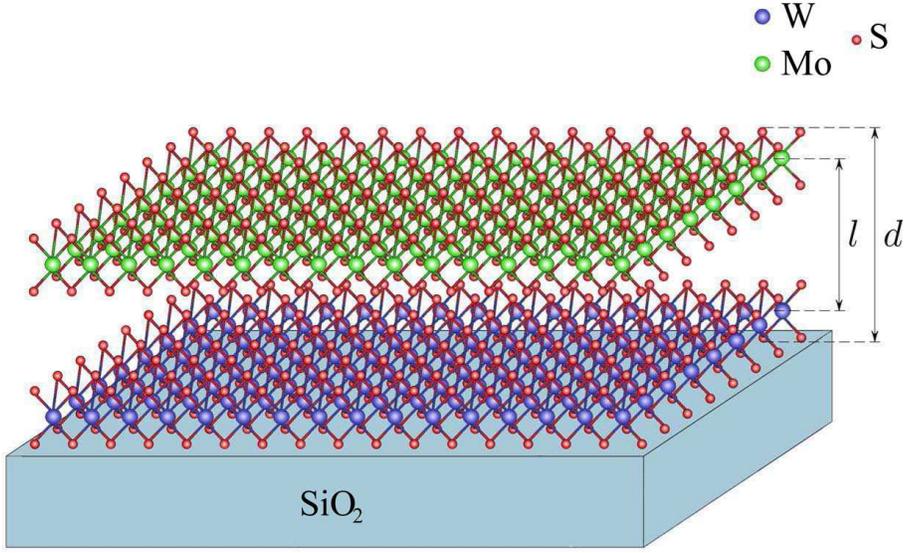}
\end{center}
\caption{The type II heterostructure 1L-WS$_2$/1L-MoS$_2$. $l$ is the characteristic distance between charged layers, $d$ is the thickness of the heterostructure.}
\label{f1}
\end{figure}

The second term is the electrostatic contribution\footnote{In view of the significant anisotropy of the permittivity in TMDs, it should be noted that $\varepsilon$ for the material of the film is understood hereinafter as the out-of-plane permittivity.}
\begin{equation}\label{E_elst}
E_\text{elst}=2\pi\frac{\varepsilon_\text{eff}}{\varepsilon}ln,
\end{equation}
where $\varepsilon$ is the static permittivity of TMD monolayers (the average of the static permittivities of each monolayer), $l$ is the characteristic distance between the layer of holes in the first material and the layer of electrons in the second material, approximately equal to the distance between adjacent planes of transition metal atoms (see Fig. \ref{f1}).

The third term is the dipole--dipole interaction contribution. By decomposing the dipole moments into parallel and perpendicular components to the $xy$ plane, it is easy to verify that the contribution of the parallel components vanishes when we average over all possible their orientations. Therefore, we can assume that we have two dipoles parallel to each other, oriented perpendicular to the heterostructure plane. They are separated by the average distance $\bar{r}=(\pi n)^{-1/2}$ between particles in EHL. The dipole interaction energy $\propto\mathbf{d}_1\mathbf{d}_2/\tilde{\varepsilon}\bar{r}^3$ includes the permittivity $\tilde{\varepsilon}$ determined by the permittivities of both the environment and the film material, $\varepsilon_\text{eff}\lesssim\tilde{\varepsilon}\lesssim\varepsilon$. There are two limiting cases: $\tilde{\varepsilon}=\varepsilon$ when two point dipoles are at a small distance from each other, $\bar{r}\ll d$, and $\tilde{\varepsilon}=\varepsilon_\text{eff}$ when the distance between them is large, $\bar{r}\gg d$. In the case of TMD bilayers and characteristic EHL densities, we are in an intermediate region, but closer to the second limit\footnote{In addition, the transition to this limit is carried out fairly quickly, if we take into account that the electric field lines emerge perpendicular to the monolayers, and the charges that make up the dipoles are located near the film surface; therefore, to pass to the second limit, it suffices $\bar{r}\gtrsim d$, which is just what is done for the system under consideration.}. To simplify the notation, we take $\tilde{\varepsilon}\approx\varepsilon_\text{eff}$. We also accept that due to the repulsion between the dipoles, they tend to be ordered into a triangular lattice. Summing up the relative contributions from the nearest neighbors, we obtain an additional coefficient. We estimate the contribution of the next nearest neighbors as $1/2^3$ from the contribution of the nearest neighbors, because the average distance has doubled. Accordingly, the contribution of the next neighbors can be estimated as $1/3^3$ from the contribution of the contribution of the nearest neighbors (the average distance has tripled), and so on, i.e. resulting in the sum $\sum\limits_{n=1}^\infty n^{-3}=\zeta(3)$, where $\zeta$ is the Riemann zeta function, $\zeta(3)\approx1.202$. Then we find
\begin{equation}\label{E_dd}
E_\text{dd}\approx5\left(\frac{5}{8}+\frac{1}{9\sqrt{3}}\right)\zeta(3)\pi^{3/2}l^2n^{3/2}.
\end{equation}

The fourth term in \eqref{E_gs} is the exchange energy \cite{Pekh2021}
\begin{equation}\label{E_exch}
E_\text{exch}=-\frac{4\sqrt{2}}{3\pi}\left(1+\sqrt{\varkappa}\right)r^{-1}_s.
\end{equation}

The last term is the correlation energy, represented as the integral over the transferred momentum \cite{Combescot1972, Andryushin1976a, Andryushin1976b, Andryushin1977}
\begin{equation}\label{E_corr}
E_\text{corr}=\int\limits_0^\infty I(q)dq.
\end{equation}

The calculated value of the exciton Bohr radius $a_B$ in TMD monolayers is several times larger than $a_x$ due to the fact that the Keldysh potential \cite{Rytova1967, Keldysh1979} ($r_0$ is the ``screening length'')
\begin{equation}\label{Keldysh_potential}
V_\text{K}(q)=\frac{2\pi\widetilde{e}^2}{q(1+r_0q)}
\end{equation}
is weaker than the usual 2D Coulomb potential $V_\text{2D}(q)=2\pi\widetilde{e}^2/q$. The value of $l\simeq6$~\AA~is always less than $a_B\gtrsim10$ \AA~[this is shown in the next section]. In this case, to take into account the correlations of electrons and holes, it is not essential whether they are separated over monolayers or are located in one monolayer. Therefore, one can use the expression for the function $I(q)$ obtained in the absence of the $e$-$h$ separation.

The function $I(q)$ is known in a small and large $q$ limits. To find $I(q)$ for $q\ll1$, we use the random phase approximation (RPA); for $q\gg1$, it is determined by the sum of second-order interaction diagrams \cite{Combescot1972, Andryushin1976a, Andryushin1976b, Andryushin1977}
\begin{equation}\label{Iq}
I(q)=\begin{cases}
-aq+bq^{3/2}-cq^2+dq^{5/2}+fq^3, & q\ll1,\\
-gq^{-3},& q\gg1,
\end{cases}
\end{equation}
where
\begin{equation*}
\begin{split}
a&=\frac{\sqrt{2}}{\pi r_s}\left(1+\frac{1}{\sqrt{\varkappa}}\right),~b=\frac{2^{1/4}}{r^{3/2}_s\nu^{1/2}_e},\\
c&=\frac{2}{\pi r^2_s\nu_e\eta_e}\left[\pi\Theta(\varkappa,\,\sigma)-f_1(\varkappa,\,\sigma)-f_2(\varkappa,\,\sigma)\right],~d=\frac{3(\eta^{-3}_e+\varkappa\eta^{-3}_h)}{2^{5/4}r^{5/2}_s\nu^{3/2}_e},\\
f=&\frac{1}{\sqrt{2}\pi r^3_s}\left\{\frac{r^2_s}{12}\left(1+\varkappa^{-3/2}\right)
-\frac{4}{\nu_e\nu_h\eta_e\eta_h}\left[\frac{f_3(\varkappa,\,\sigma)}{2(1+\sigma\varkappa)^2}+f_4(\varkappa,\,\sigma)\right]\right\},\\
g&=\nu_e\left(\eta_e+\eta_h+4\right)-\frac{\eta_e}{2}-\frac{\varkappa\eta_h}{2},~\eta_e=\frac{m_{e2}}{m}=1+\sigma,~\eta_h=\frac{m_{h1}}{m}=1+\sigma^{-1},
\end{split}
\end{equation*}
\begin{equation*}
\begin{split}
f_1(\varkappa,\,\sigma)&=\int\limits_0^1\arctan\left[\frac{x}{1+\varkappa\sigma}\left(\frac{\varkappa}{\sqrt{1-x^2}}+\frac{1}{\sqrt{\varkappa\sigma^2-x^2}}\right)\right]dx,\\
f_2(\varkappa,\,\sigma)&=\int\limits_1^{\sigma\sqrt{\varkappa}}\arctan\left[\frac{x}{\sqrt{\varkappa\sigma^2-x^2}\left(1+\varkappa\sigma-\frac{\varkappa\sigma x}{\sqrt{x^2-1}}\right)}\right]dx,\\
f_3(\varkappa,\,\sigma)&=\int\limits_0^1\frac{\varkappa\sigma/\sqrt{1-x}+1/\sqrt{\varkappa\sigma^2-x}}{1+\frac{x}{(1+\varkappa\sigma)^2}
\left[\frac{\varkappa\sigma}{\sqrt{1-x}}+\frac{1}{\sqrt{\varkappa\sigma^2-x}}\right]^2}dx,\\
f_4(\varkappa,\,\sigma)&=\int\limits_1^{\sigma\sqrt{\varkappa}}\frac{x\sqrt{\varkappa\sigma^2-x^2}dx}{\left(\varkappa\sigma^2-x^2\right)\left[1+\varkappa\sigma-\frac{\varkappa\sigma x}{\sqrt{x^2-1}}\right]^2+x^2},\\
\Theta(\varkappa,\,\sigma)&=1+\left(\sigma\sqrt{\varkappa}-\frac{1+\sigma\varkappa}{\sqrt{1+2\sigma\varkappa}}\right)
\theta\left(\varkappa-\frac{2-\sigma+\sqrt{4\sigma+\sigma^2}}{2\sigma(2\sigma-1)}\right),
\end{split}
\end{equation*}
$\theta(x)$ is the Heaviside function; it is assumed that $\sigma\sqrt{\varkappa}>1$.

We have given formulas for the mass ratio $\sigma$, which can be different from 1 (often, $\sigma\gtrsim1$ as, for example, in 1L-WS$_2$/1L-MoS$_2$ and 1L-WSe$_2$/1L-MoSe$_2$). Due to the presence of the electrostatic contribution in $E_\text{gs}$, the minimum of the EHL energy can be very close to the exciton energy. We present formulas with $\sigma\neq1$ for a more accurate calculation of the EHL binding energy.

In the intermediate region of $q$, the function $I(q)$ is approximated by a segment of the common tangent of both asymptotics \eqref{Iq}. Integrating $I(q)$ over $q$, we find
\begin{equation}\label{E_corr1}
\begin{split}
&E_\text{corr}=\frac{1}{2}\left(-aq_1+bq^{3/2}_1-cq^2_1\right)q_2+\frac{1}{2}\left(dq_2-\frac{b}{5}\right)q^{5/2}_1\\
&+\frac{1}{2}\left(fq_2+\frac{c}{3}\right)q^3_1
-\frac{3d}{14}q^{7/2}_1-\frac{f}{4}q^4_1-\frac{g}{q^2_2}\left(1-\frac{q_1}{2q_2}\right),
\end{split}
\end{equation}
where $q_1$ and $q_2$ are the matching points of the asymptotics \eqref{Iq} with the line segment.

The point $q_1$ lies near the point $q_0$ of the minimum of the $I(q)$ asymptotics for $q\ll1$. The point $q_0$ is near 1 for not too large $\nu_{e,h}$ and $1\lesssim r_s\lesssim2$
\begin{equation*}
q_1\approx q_0\approx\frac{a-\frac{3}{4}b+\frac{5}{4}d+3f}{\frac{3}{4}b-2c+\frac{15}{4}d+6f}.
\end{equation*}

We find for $\nu_{e,h}\gg1$
\begin{equation*}
q_1\approx q_0\approx2\sqrt{2}\sqrt{\frac{\varkappa+\varkappa^{3/2}}{1+\varkappa^{3/2}}}.
\end{equation*}

With a good accuracy (within 10\%) the second matching point is
\begin{equation*}
q_2\approx\left(\frac{4g}{|I(q_0)|}\right)^{1/3}.
\end{equation*}

In the multivalley case ($\nu_{e,h}\gg1$), we obtain the lower bound for the correlation energy
\begin{equation}\label{Ecorr2}
E_\text{corr}\gtrsim-A(\varkappa,\,\sigma)n^{1/3},
\end{equation}
where
\begin{equation*}
A(\varkappa,\,\sigma)=\left(\frac{12}{\pi}\right)^{1/3}\frac{\varkappa^{1/6}\left(1+\sqrt{\varkappa}\right)^{2/3}}{\left(1+\varkappa^{3/2}\right)^{1/3}}\left(1+\sigma+\sigma^{-1}\right)^{1/3}.
\end{equation*}

To compare the EHL ground state energy with the exciton binding energy, it is necessary to know the latter. It is also necessary to evaluate the influence of the separation of an electron and a hole over different monolayers on the exciton binding energy. This is the subject of the next section.

\section{Binding energy of the interlayer exciton}
To begin with, we get the potential created by the charge in one monolayer at a point located inside the other monolayer.

Let there be a dielectric medium (substrate) with permittivity $\varepsilon_1$ below the film, and a medium with permittivity $\varepsilon_2$ located above it. We describe a bilayer film as composed of films of equal thickness (in $d/2$) with permittivities $\varepsilon^{(1)}$ and $\varepsilon^{(2)}$ of the materials of the first and second monolayers, respectively, $\varepsilon^{(1)}\approx\varepsilon^{(2)}$. We presume\footnote{To obtain the general expression \eqref{phi1}, it suffices $\varepsilon^{(1,2)}>\varepsilon_{1,2}$.} $\varepsilon^{(1,2)}\gg\varepsilon_{1,2}$.

From the point of view of electrodynamics of continuous media, there is no gap between monolayers, since each monolayer is considered as a continuous medium and the bilayer consists of two tightly contacting layers with a common boundary in the middle of the film.

Following works \cite{Rytova1967, Keldysh1979}, it is not difficult to obtain the potential created at a point $\mathbf{r}=(\boldsymbol{\rho},\,z)$ by a charge at a point $\mathbf{r}^\prime=(\mathbf{0},\,z^\prime)$ (these two points are in different monolayers)
\begin{equation}\label{phi1}
\varphi(\rho,\,z,\,z^\prime)=2e\int\limits_0^\infty\frac{\cosh\left(k\left(\frac{d}{2}+z^\prime\right)+\eta_1\right)\cosh\left(k\left(\frac{d}{2}-z\right)+\eta_2\right)J_0(k\rho)dk}
{\varepsilon^{(1)}\sinh\left(k\frac{d}{2}+\eta_1\right)\cosh\left(k\frac{d}{2}+\eta_2\right)+\varepsilon^{(2)}\cosh\left(k\frac{d}{2}+\eta_1\right)\sinh\left(k\frac{d}{2}+\eta_2\right)},
\end{equation}
where $J_0$ is the Bessel function of the first kind and $\eta_{1,2}=\artanh\left(\varepsilon_{1,2}/\varepsilon^{(1,2)}\right)$.

Essential $k$ at $\rho\gg d$ are such that $kd\ll1$ \cite{Keldysh1979}, therefore, in the linear approximation with respect to $k$, $J_0(k\rho)$ remains in the numerator of the integrand of \eqref{phi1}, and the denominator, taking into account $\eta_{1,2}\approx\varepsilon_{1,2}/\varepsilon^{(1,2)}$ at $\varepsilon^{(1,2)}\gg\varepsilon_{1,2}$, has $\varepsilon^{(1)}\left(k\frac{d}{2}+\eta_1\right)+\varepsilon^{(2)}\left(k\frac{d}{2}+\eta_2\right)\approx\overline{\varepsilon}d\left[k+r^{-1}_0\right]$ with $\overline{\varepsilon}=\left(\varepsilon^{(1)}+\varepsilon^{(2)}\right)/2$ and $r_0=\overline{\varepsilon}d/\left(\varepsilon_1+\varepsilon_2\right)$. Then the potential does not depend on $z$ and $z^\prime$ and is equal to
\begin{equation}\label{phi2}
\varphi(\rho)=\frac{\pi e}{2\varepsilon_\text{eff}r_0}\left[H_0\left(\frac{\rho}{r_0}\right)-Y_0\left(\frac{\rho}{r_0}\right)\right],
\end{equation}
where $H_0$ and $Y_0$ are the Struve function and the Bessel function of the second kind and, as above, $\varepsilon_\text{eff}=\left(\varepsilon_1+\varepsilon_2\right)/2$.

We obtained the potential that coincides with the Keldysh potential as a function of $\rho$, but with a different value of $r_0$. The appearance of $\overline{\varepsilon}$ as a half-sum of $\varepsilon^{(1)}$ and $\varepsilon^{(2)}$ is a direct consequence of the model statement about equal thicknesses of the layers that make up the film. Otherwise, $\varepsilon^{(1)}$ and $\varepsilon^{(2)}$ would be included with the proportions $d_1/d$ and $d_2/d$, $d=d_1+d_2$, $d_{1,2}$ are the layer thicknesses. We have $d_1\approx d_2$ (within a few tenths of \AA) for the film of two monolayers and taking into account the difference between them would mean a clear overestimation of the accuracy. For a monolayer/bilayer film, we have $d_2\approx2d_1$ and $\overline{\varepsilon}=\frac{1}{3}\varepsilon^{(1)}+\frac{2}{3}\varepsilon^{(2)}$.

In the case of small $\rho$, it should be taken into account that an electron and a hole separated over monolayers cannot be arbitrarily close to each other (the minimum distance between them is $\approx l$). On the other hand, to solve the problem about 2D exciton, the obtained potential should be averaged over the $z$-coordinates of the electron and hole.

Let there be a hole (a charge that creates the potential) in the lower monolayer, $-\frac{d}{2}\leq z^\prime\leq\frac{d}{2}-l$, and an electron in the upper monolayer, $l-\frac{d}{2}\leq z\leq\frac{d}{2}$. To simplify the consideration, we consider each monolayer in the transverse direction as a potential well with infinitely high walls. The wave functions of ``transverse motion'' are $\psi_1(z^\prime)=\sqrt{\frac{2}{d-l}}\cos\left(\pi\frac{z^\prime+l/2}{d-l}\right)$ and $\psi_2(z)=\sqrt{\frac{2}{d-l}}\cos\left(\pi\frac{z-l/2}{d-l}\right)$. The average potential is
\begin{equation}\label{phi3}
\overline{\varphi}(\rho)=\int\limits_{-d/2}^{d/2-l}dz^\prime\int\limits_{l-d/2}^{d/2}dz\varphi(\rho,\,z,\,z^\prime)
\left|\psi_1(z^\prime)\right|^2\left|\psi_2(z)\right|^2=2^7\pi^4e\int\limits_0^\infty\frac{\mathcal{F}(k)J_0(k\rho)dk}{k^2(d-l)^2\left(k^2(d-l)^2+4\pi^2\right)^2},
\end{equation}
where
\begin{equation*}
\mathcal{F}(k)=\frac{\cosh\left(\frac{1}{2}k(d-l)+\eta_1\right)\cosh\left(\frac{1}{2}k(d-l)+\eta_2\right)\sinh^2\left(\frac{1}{2}k(d-l)\right)}
{\varepsilon^{(1)}\sinh\left(k\frac{d}{2}+\eta_1\right)\cosh\left(k\frac{d}{2}+\eta_2\right)+\varepsilon^{(2)}\cosh\left(k\frac{d}{2}+\eta_1\right)\sinh\left(k\frac{d}{2}+\eta_2\right)}.
\end{equation*}

For simplicity, we neglect the shift of the maximum of the wave function of electrons and holes towards the vdW gap due to the Coulomb attraction between them. Below we show the smallness of the effect of electron (hole) density smearing across monolayers. The density shift is an effect of a higher order of smallness (an additional smallness parameter $(d-l)/2d\approx\hspace{0.05cm}^1\hspace{-0.08cm}/_6$ arises owing to the geometry of the system). Physically, the shift in the transverse direction of the electron density due to an attraction of the hole is still small at the average distance in the monolayer plane between the electron and hole $\langle\rho\rangle=a_B\gtrsim d$.

Having choused potential wells with infinitely high walls, we neglected tunneling between monolayers (otherwise the wave functions $\psi_1(z^\prime)$ and $\psi_2(z)$ would overlap and the tunneling probability would be nonzero). Tunneling is important, in particular, for estimating the charge carrier lifetimes. We consider the interlayer exciton as a single bound state, when there is only one electron and one hole. So, tunneling currents between monolayers can be neglected.

Hamiltonian of the interlayer exciton is
\begin{equation}\label{Ham_2Dexciton}
\widehat{H}_\text{ex}=-\frac{1}{2m}\triangle_{\boldsymbol\rho}+V(\rho),
\end{equation}
where $\triangle_{\boldsymbol\rho}$ is the Laplacian in the polar coordinates, $V(\rho)=-e\overline{\varphi}(\rho)$ is the interaction energy of the electron and hole, $m$ is their reduced mass.

To find the binding energy of the interlayer exciton, we use the variational method. The trial wave function of the 1$s$ state is chosen in the form of the exponentially decreasing function \cite{Ratnikov2020}
\begin{equation}\label{trial_wave_fun}
\widetilde{\psi}(\rho)=\sqrt{\frac{2}{\pi}}a^{-1}e^{-\rho/a},
\end{equation}
where $a$ is the variational parameter.

\begin{figure}[t!]
\begin{center}
\includegraphics[width=12cm]{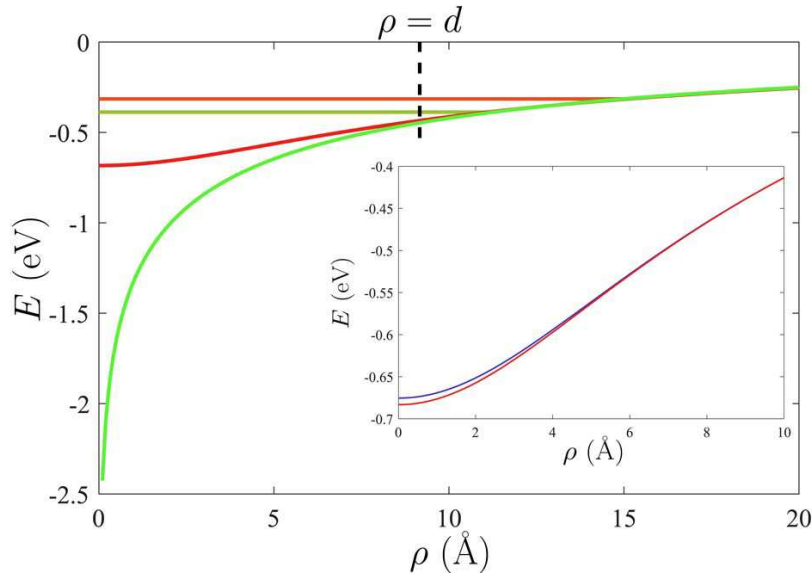}
\end{center}
\caption{The potentials \eqref{phi3} (the red curve) and \eqref{phi2} (the green curve) for the heterostructure 1L-WS$_2$/1L-MoS$_2$ on the SiO$_2$ substrate. The red and dark green segments show the exciton levels in the ground state corresponding to these two potentials, $-315$~meV and $-388$~meV. The Bohr radii are 12.4~\AA~and 9.7 \AA, respectively. The inset shows two potential curves with allowance for density smearing (red curve) and for the fixed distance $l$ between the electron and hole planes (blue curve). The lattice constants along the $c$-axis are taken from \cite{Wilson1969}, $l=(c_1+c_2)/4=6.164$~\AA. The averaged thickness of the monolayer $d-l=3$~\AA~is taken from \cite{Py1983}. The out-of-plane permittivities are taken from~\cite{Laturia2018}.}
\label{f2}
\end{figure}

The energy of the interlayer exciton in units of $E_x$ is
\begin{equation}\label{E_ex}
E_\text{ex}=a^{-2}_0-2^7\pi^4\varepsilon_\text{eff}\int\limits_0^\infty\frac{\mathcal{F}(k)dk}
{k^2(d-l)^2\left(k^2(d-l)^2+4\pi^2\right)^2\left(1+\frac{1}{4}a^2_0k^2\right)^{3/2}}.
\end{equation}
Here, $a_0$, $d$ and $l$ are measured in units of $a_x$, $k$ is measured in units of~$a^{-1}_x$. $a_0$ is the value of the variational parameter corresponding to the energy minimum; it satisfies the equation
\begin{equation}\label{a0}
a^{-4}_0-\frac{3(2\pi)^4\varepsilon_\text{eff}}{(d-l)^2}\int\limits_0^\infty\frac{\mathcal{F}(k)dk}
{\left(k^2(d-l)^2+4\pi^2\right)^2\left(1+\frac{1}{4}a^2_0k^2\right)^{5/2}}=0.
\end{equation}

We note that the potential \eqref{phi3} is finite as $\rho\rightarrow0$, in contrast to the potential \eqref{phi2}, which, as is well known, goes to infinity logarithmically. Hence it follows that the binding energy of the 2D exciton with Hamiltonian \eqref{Ham_2Dexciton}, where potential \eqref{phi2} is substituted, is always greater than $|E_\text{ex}|$ found by formula \eqref{E_ex}. Figure \ref{f2} shows the typical dependencies $-e\varphi(\rho)$ and $-e\overline{\varphi}(\rho)$ and the position of the exciton energy in the ground state, obtained by the variational calculation. At $\rho\gtrsim d$, these two potentials practically merge. The binding energies differ within 20\%. The inset also shows the dependencies $-e\overline{\varphi}(\rho)$ and $-e\varphi(\rho,\,l/2,\,-l/2)$ [according to formula \eqref{phi1}], demonstrating the smallness of the effect of electron (hole) density smearing across monolayers: these two curves practically merge at $\rho\gtrsim d/2$, while the difference at $\rho\rightarrow0$ is only $\approx1$\%.

We see that the assumed model positions are confirmed by calculations. The weak change in the exciton binding energy for two potentials (with finite depth and decreasing logarithmically) is due to a rather weak tendency to $-\infty$ compared to $-1/\rho$.

\section{Numerical calculations: EHL vs exciton}

To demonstrate the possibility of EHL formation, let us compare the energies of the EHL and exciton ground state for systems of two variants: monolayer/monolayer and monolayer/bilayer. The distances between the layers of electrons and holes in both cases are considered the same (in the monolayer/bilayer system, the electrons in the bilayer, being attracted to the holes in the monolayer, will move closer to the holes, as in the monolayer/monolayer system). The number of electron valleys in the monolayer/bilayer system was taken into account as 3 (spin degeneracy at the $\Lambda$ points was removed). The effective masses of electrons in the valleys at the $\Lambda$ points were assumed to be equal for monolayers and bilayers (we do not have data on the effective masses of electrons in the valleys of the $\Lambda$ points in bilayers).

The parameters required for the calculation are presented in Table~\ref{t1}. $m_0$ is the free electron mass. The values separated by commas indicate the parameters for the monolayer/monolayer and monolayer/bilayer variants of the heterostructures, respectively. The dielectric environment of the systems is the silicon dioxide substrate ($\varepsilon_1=3.9$) and vacuum ($\varepsilon_2=1$).

We use the previously approved method \cite{Pekh2021}, when we move away from the Coulomb units of the 2D exciton for the EHL calculation. The value of length dimension $l$ are dimensionless by the calculated Bohr radius $a_B$. To convert the density into cm$^{-2}$, we take $a^{-2}_B$ as the unit of measurement. The energy is measured in units of the calculated exciton binding energy $|E_\text{ex}|$.

\begin{table}[b!]
\caption{\label{t1} Parameters of systems for EHL and exciton calculations, the Bohr radius $a_B$, exciton binding energy $|E_\text{ex}|$, the equilibrium density $n_0$ and binding energy $|E_0|$ of EHL.}
\begin{tabular}{|l|c|c|c|c|l|}
\hline
Parameter & WS$_2$/MoS$_2$ & WSe$_2$/MoSe$_2$ & WS$_2$/MoSe$_2$ & WSe$_2$/MoS$_2$ & Ref.\\
\hline
$l$ (\AA)                   & 6.164         & 6.474         & 6.321         & 6.318         & \cite{Wilson1969}\\
$d$ (\AA)                   & 9.164, 15.328 & 9.579, 16.053 & 9.366, 15.687 & 9.378, 15.696 & \cite{Py1983}\\
$\overline{\varepsilon}$    & 6.35, 6.63    & 7.45, 7.77    & 6.85, 7.37    & 6.95, 7.03    & \cite{Laturia2018}\\
$m_{h1}$ ($m_0$)            & 0.42          & 0.44          & 0.42          & 0.44          & \cite{Rasmussen2015}\\
$m_{e2}$ ($m_0$)            & 0.67, 0.81    & 0.49, 0.71    & 0.49, 0.71    & 0.67, 0.81    & \cite{Eknapakul2014}, \cite{Selig2018}\\
\hline
$a_B$ (\AA)                 & 12.41, 13.96  & 13.99, 15.19  & 13.70, 14.98  & 12.68, 14.18 & this\\
$|E_\text{ex}|$ (meV)       & 315, 244      & 274, 216      & 287, 223      & 300, 235     & work\\
$n_0$ (10$^{12}$ cm$^{-2}$) & 4.5, 3.6      & 3.8, 4.1      & 3.9, 4.2      & 4.3, 4.4     & \\
$|E_0|$ (meV)               & 307, 243      & 270, 239      & 282, 248      & 293, 258     & \\
\hline
\end{tabular}
\end{table}

\begin{figure}[t!]
\begin{center}
\includegraphics[width=\textwidth]{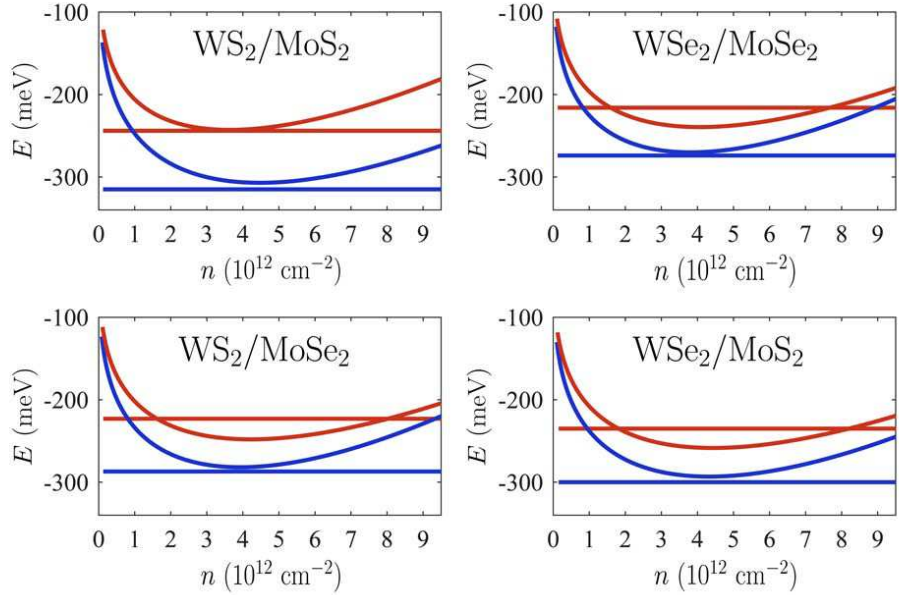}
\end{center}
\caption{Numerical calculation of dependence of EHL ground state energy on density. The blue curves correspond to the monolayer/monolayer heterostructures, and the red curves correspond to the monolayer/bilayer heterostructures. The horizontal segments of the corresponding colors show the positions of the exciton levels.}
\label{f3}
\end{figure}

Figure \ref{f3} shows the numerical calculation results of the dependence of the EHL energy on density for four systems with the type II contact, each of which is presented in two variants, monolayer/monolayer and monolayer/bilayer. For comparison, the positions of the exciton levels are also shown as horizontal segments. According to our calculations, EHL in the first heterostructure is still not energetically favorable in comparison with the exciton in both variants. In other heterostructures, the EHL formation is possible: for the first variant of the heterostructures, the minimum energy of the EHL ground state lies above the exciton level, and for the second variant, it lies below the corresponding exciton level. The salience of the first heterostructure is explained apparently by the lowest permittivity $\overline{\varepsilon}$ of the presented four.

The physical reason for this behavior lies in the fact that, on the one hand, the contribution of the kinetic energy in $E_\text{gs}$ decreases with an increase in the number of valleys, and, on the other hand, an increase in the film thickness leads to a weakening of the Coulomb interaction. The exciton energy changes more strongly than the EHL ground state energy, because in the case of EHL this leads to multidirectional trends (the $E_\text{elst}$ and $E_\text{dd}$ contributions decreases due to an increase in $a_B$ and this leads to a decrease in the energy, and both the exchange and correlation energies in modulus decreases and this leads to a increase in the energy).

\section{Gas--liquid transition}
In 1968 Keldysh suggested \cite{Keldysh1968} that as the temperature decreases or the density of exciton or exciton molecule (biexciton) gas increases, this gas will liquefy in EHL. The $e$-$h$ system would undergo a first-order gas--liquid phase transition. EHL was discovered as early as next year in germanium crystals \cite{Pokrovskii1969, Asnin1969, Vavilov1969, Bagaev1969}. Since then, EHL has also been observed in Si, GaP, GaAs, CdS and many other semiconductors (see, for example, \cite{Rice1977, Hensel1977, Jeffries1983, Tikhodeev1985, Sibeldin2016, Sibeldin2017} and references therein).

The gas--liquid transition in $(n,\,T)$ plane corresponds to a bell-shaped curve with the critical point $(n_c,\,T_c)$ at the top. This curve passes at $T=0$ through the points $n=\mathfrak{n}$ and $n=n_0$. The value of $\mathfrak{n}$ is determined by the equality $|E_0|=|E_\text{ex}|$ (biexcitons most likely do not exist due to dipole-dipole repulsion in a system with separated electrons and holes; otherwise, half of the biexciton dissociation energy should be added to $|E_\text{ex}|$). The region to the left of the branch $n_\text{G}$ emerging from the point $(\mathfrak{n},\,0)$ corresponds to the exciton gas. The region to the right of the branch $n_\text{L}$ leaving the point $(n_0,\,0)$ is occupied by EHL. Below the bell-shaped curve is the region of coexistence of EHL droplets with the density $n_\text{L}$ and a saturated vapor of excitons or $e$-$h$ plasma with the density $n_\text{G}$.

We represent the chemical potential of $e$-$h$ pair ensemble in type II TMD heterostructures as
\begin{equation}\label{mu}
\mu(n,\,T)=T\ln\left[\left(e^{\frac{2\pi n}{(1+\sigma)\nu_eT}}-1\right)\left(e^{\frac{2\pi\sigma n}{(1+\sigma)\nu_hT}}-1\right)\right]+\frac{\partial\Delta F}{\partial n},
\end{equation}
where the first term is the chemical potential of noninteracting particles (the kinetic contribution), and $\Delta F$ is the contribution to the free energy from all other (besides $E_\text{kin}$) terms of the ground state energy, $\Delta F=n\left(E_\text{elst}+E_\text{dd}+E_\text{exch}+E_\text{corr}\right)$. The temperature dependence enters only through the kinetic contribution, since the temperature corrections to the remaining terms either cancel each other out, as in the case of the exchange and correlation contributions \cite{Andryushin1979}, or are small in the parameter $T/E_F$ ($E_F=q^2_{Fe}/2m_{e2}+q^2_{Fh}/2m_{h1}$ is the Fermi energy of $e$-$h$ pairs).

The critical point is determined by two equations
\begin{equation}\label{critpoint}
\left.\frac{\partial\mu}{\partial n}\right|_{\substack{n=n_c\\ T=T_c}}=\left.\frac{\partial^2\mu}{\partial n^2}\right|_{\substack{n=n_c\\ T=T_c}}=0.
\end{equation}

Previously, we numerically verified for 1L-TMD heterostructures \cite{Pekh2021} the characteristic relationship for the gas--liquid transition between the critical temperature and the binding energy of the liquid $T_c\simeq0.1|E_0|$ \cite{Landau5}. We obtained for the considered heterostructures $T_c\approx\frac{1}{8}|E_0|$ and $n_c\approx\frac{1}{5}n_0$.

Here, we again check the indicated relationship already for the monolayer/bilayer type II TMD heterostructures. The results are presented in Table~\ref{t2}. We have $T_c\approx\frac{1}{9}|E_0|$ and $n_c\approx\frac{1}{6}n_0$. Both of these ratios decrease: the ratio of densities becomes even more different from the value in three-dimensional semiconductors ($n_c/n_0\approx\frac{1}{3}$), while the ratio of critical temperature to binding energy becomes closer to the corresponding value in them ($T_c/|E_0|\approx\frac{1}{10}$) \cite{Tikhodeev1985}.

The shift of the critical point to the region of lower densities and temperatures compared to monolayer heterostructures is explained by the presence of contributions $E_\text{elst}$ and $E_\text{dd}$ in the EHL ground state energy. Due to their positiveness, lower densities and temperatures are required for a liquid decomposition.

\begin{table}[b!]
\caption{\label{t2} The critical point of the gas--liquid transition and the ratio of the equilibrium parameters to the critical ones for the EHL in the monolayer/bilayer type II TMD heterostructures.}
\begin{center}
\begin{tabular}{|l|c|c|c|}
\hline
Parameter & WSe$_2$/MoSe$_2$ & WS$_2$/MoSe$_2$ & WSe$_2$/MoS$_2$\\
\hline
$n_c$ (10$^{11}$ cm$^{-2}$) & 6.4  & 6.7  & 6.9 \\
$T_c$ (K)                   & 311  & 321  & 334 \\
$n_0/n_c$                   & 6.32 & 6.34 & 6.40\\
$|E_0|/T_c$                 & 8.95 & 8.95 & 8.97\\
\hline
\end{tabular}
\end{center}
\footnotesize{Note: $T_c$ in the last line is measured in meV.}
\end{table}

To obtain the temperature dependence of densities $n_\text{G}(T)$ and $n_\text{L}(T)$ of $e$-$h$ pairs in the gas and liquid phases, we use the Maxwell construction
\begin{equation}\label{Maxwell_constrution}
\int\limits_{n_\text{G}}^{n_\text{L}}\mu(n,\,T)dn=\mu(T)(n_\text{L}-n_\text{G}),
\end{equation}
where $\mu(T)=\mu(n_\text{G},\,T)=\mu(n_\text{L},\,T)$.

\begin{figure}[t!]
\begin{center}
\includegraphics[width=0.85\textwidth]{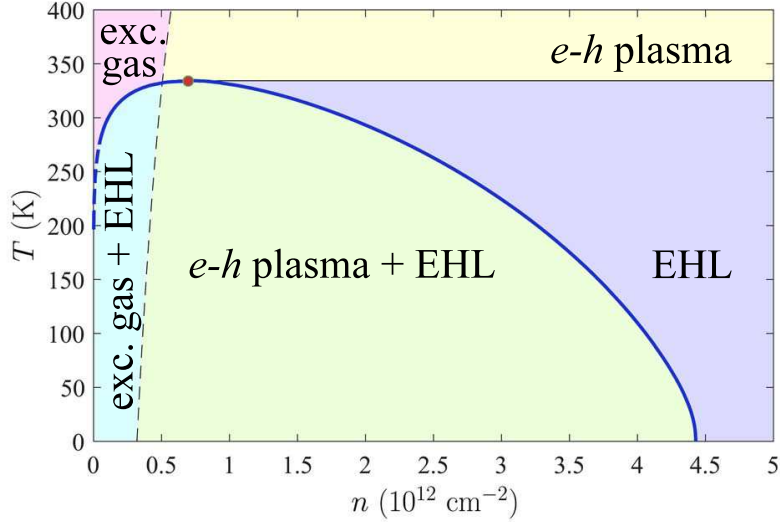}
\end{center}
\caption{Phase diagram for the transitions between an exciton gas, $e$-$h$ plasma, and EHL in the 1L-WSe$_2$/2L-MoS$_2$ heterostructure. The red circle indicates the critical point of the gas--liquid transition. The dashed segment of the coexistence curve indicates the region of low charge carrier densities where RPA becomes inapplicable.}
\label{f4}
\end{figure}

The set of pairs of points $n_\text{G}$ and $n_\text{L}$ forms a bell-shaped curve in the plane $(n,\,T)$, also called a coexistence curve. The typical coexistence curve is presented in Figure~\ref{f4}.

There is also the metal--insulator transition. The temperature dependence of the metal--insulator transition density $n_{dm}(T)$ is shown in Figure~\ref{f4} by the black dotted line. As a rule, the curve $n_{dm}(T)$ crosses the gaseous branch $n_\text{G}(T)$ of the coexistence curve (the metal--insulator transition occurs in the gas phase). Above the coexistence curve, the curve $n_{dm}(T)$ is the boundary between the region occupied by the exciton gas and the region containing the $e$-$h$ plasma. We plotted $n_{dm}(T)$ here qualitatively based on our previous calculations of monolayer heterostructures \cite{Pekh2021}. An accurate consideration of the metal--insulator transition in the monolayer/bilayer TMD type II heterostructures is beyond the scope of this article and will be carried out by us in another work.

\section{EHL line in the photoluminescence spectrum}

It is well known that the EHL line in the photoluminescence spectrum extends from the narrow exciton line towards lower energies by the value of the exciton work function from EHL $\phi=E_\text{ex}-E_0$. At low temperatures, the EHL line width is equal to the Fermi energy of $e$-$h$ pairs $E_F$. However, calculating the EHL line shape turns out to be a rather difficult problem.

To begin with, we note that electrons in the monolayer/bilayer type~II TMD heterostructures are located in the valleys at the $\Lambda$ points, while the holes remain in the valleys at the $K$ points. This means that optical transitions are indirect in the ${\bf k}$ space and requires the participation of phonons in the wave vectors about a quarter of the size of the Brillouin zone \cite{Splendiani2010}. The well-developed method for processing the EHL line shape in indirect semiconductors is quite applicable to describe the process of the EHL recombination in the systems under consideration. The dependence of intensity on frequency is determined by the expression \cite{Sibeldin2016, Sibeldin2017}
\begin{equation}\label{Line_shape_general}
I(\omega)\propto\int\int D_e(E_e)D_h(E_h)f_e(E_e)f_h(E_h)\delta\left(E_e+E_h+E_{g\text{L}}-\Omega-\omega\right)dE_edE_h,
\end{equation}
where $D_e$ and $D_h$ are densities of states in the conduction and valence bands, respectively, $f_e$ and $f_h$ are the Fermi distribution functions of electrons and holes with the electron and hole energies $E_e$ and $E_h$, $E_{g\text{L}}$ is the renormalized band gap in the region of the crystal occupied by EHL, $E_{g\text{L}}=E^{(0)}_g+E_0-E_F$, and $\Omega$ is the frequency of the phonon emitted in the electron transition.

We adopt an idealized model of parabolic bands in the vicinity of the corresponding valleys. The 2D nature of charge carriers greatly simplifies analytical calculations. We obtain
\begin{equation}\label{Line_shape}
\begin{split}
I(\omega)\propto&\frac{\nu_e\nu_hm_em_h}{\pi^2}\frac{\left(e^{\frac{2\pi n_\text{L}}{(1+\sigma)\nu_eT}}-1\right)\left(e^{\frac{2\pi\sigma n_\text{L}}{(1+\sigma)\nu_hT}}-1\right)\theta\left(\Omega+\omega-E_{g\text{L}}\right)}
{e^{(\Omega+\omega-E_{g\text{L}})/T}-\left(e^{\frac{2\pi n_\text{L}}{(1+\sigma)\nu_eT}}-1\right)\left(e^{\frac{2\pi\sigma n_\text{L}}{(1+\sigma)\nu_hT}}-1\right)}\\
\times&\left\{T\ln\left[\left(e^{(\Omega+\omega-E_{g\text{L}})/T}+e^{\frac{2\pi n_\text{L}}{(1+\sigma)\nu_eT}}-1\right)\left(e^{(\Omega+\omega-E_{g\text{L}})/T}+e^{\frac{2\pi\sigma n_\text{L}}{(1+\sigma)\nu_heT}}-1\right)\right]\right.\\
&\left.-\frac{\nu_h+\sigma\nu_e}{1+\sigma}\frac{2\pi n_\text{L}}{\nu_e\nu_h}-\omega-\Omega+E_{g\text{L}}\right\}.
\end{split}
\end{equation}

\begin{figure}[b!]
\begin{center}
\includegraphics[width=\textwidth]{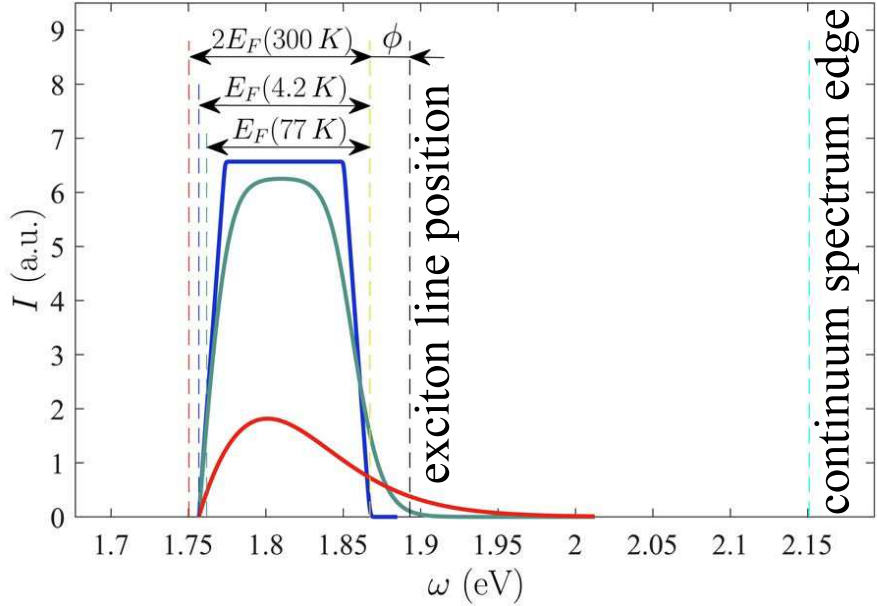}
\end{center}
\caption{EHL line in the photoluminescence spectrum of the 1L-WSe$_2$/2L-MoS$_2$ heterostructure at 4.2 K (the blue curve), 77 K (the green curve), and 300 K (the red curve).}
\label{f5}
\end{figure}

It has been demonstrated for semiconducting TMDs that the double-resonant Raman scattering has contributions from the longitudinal acoustic (LA) and transversal acoustic (TA) phonon branches \cite{Malard2013}. The Raman spectrum of TMD bilayers presents a strong intensity for the 2LA band, frequency of which is about the same between 80 and 300 K, changing just 2~cm$^{-1}$. The main contribution to the 2LA band in TMD bilayers is from the $K\Lambda$ electron-scattering process. The Raman shift for the 2LA band in 2L-MoS$_2$ is $\Delta\omega_\text{R}\approx470$ cm$^{-1}$ \cite{Gontijo2019}, i.e. $\approx58$ meV. Consequently, the frequency of the LA phonon is $\Omega=\Delta\omega_\text{R}/2\approx29$ meV. We assume below that the EHL recombination is accompanied by the emission of the LA phonon (although we do not rule out the presence of the TA component).

Figure~\ref{f5} shows three EHL line shapes for $T$\,=\,4.2, 77, and 300 K in the 1L-WSe$_2$/2L-MoS$_2$ heterostructure. The calculation was carried out according to the formula \eqref{Line_shape}. The initial (non-renormalized) gap is defined as the distance between the edges of the valence band in 1L-WSe$_2$ and the conduction band in 2L-MoS$_2$: $E^{(0)}_g=\chi\left(\text{1L-WSe}_2\right)+E_g\left(\text{1L-WSe}_2\right)-\chi\left(\text{2L-MoS}_2\right)$, where $\chi$ is the electron affinity (the position of the conduction band edge with respect to the vacuum level), $\chi\left(\text{1L-WSe}_2\right)=3.75$ eV, $\chi\left(\text{2L-MoS}_2\right)\approx\chi\left(\text{1L-MoS}_2\right)=4.1$ eV \cite{Guo2016}, $E_g\left(\text{1L-WSe}_2\right)=2.53$ eV \cite{Hanbicki2015}. The frequency of the LA phonon $\Omega$ is indicated above.

The EHL line is separated from the exciton line by the value of the exciton work function from EHL $\phi\approx25.8$ meV. At low temperatures (in particular, 4.2 and 77 K), the EHL line width is equal to the Fermi energy $E_F$ with good accuracy (the EHL density $n_\text{L}$ depends on $T$, and therefore $E_F$ is different for different temperatures). However, at high temperatures (300 K), the EHL line width turns out to be approximately twice as large as the corresponding $E_F$, which is associated with significant thermal broadening (it is comparable to $E_F$). On the whole, we see that the derived formula for the intensity correctly describes all the features of the EHL line.

\section{Discussion of the results and conclusions}
We have obtained formulas for calculating the ground state energy of EHL with separation of charge carriers over monolayers in TMD-based heterostructures. To calculate the correlation energy, one can use the same approach that was used to calculate it in the absence of separation of electrons and holes over monolayers, because the average distance between an electron and a hole $a_B$ (the Bohr radius) is always greater than the distance between charged planes $l$. For a more accurate calculation of EHL, we have given formulas for the mass ratio of charge carriers $\sigma$, which can be different from unity, and such that $\sigma\sqrt{\varkappa}>1$.

Using the variational method, we found the binding energy of the interlayer exciton. We have shown that the transverse shift of the electron and hole densities towards the vdW gap can be neglected due to the appearance of an additional small parameter. The Coulomb well has a finite depth, but the difference between the exciton level in it and the exciton level in the Keldysh potential does not exceed 20\% owing to the weak (logarithmic) tendency of the latter to $-\infty$ at $\rho\rightarrow0$.

The specificity of the problem posed on the EHL formation possibility in the type II TMD heterostructures, in particular, lies in the fact that both EHL and the exciton have to be calculated more accurately. Unfortunately, we are limited here by the approximate knowledge about the charge carrier effective masses.

It was shown that EHL becomes energetically favorable if the number of electron valleys is increased from two (with spin degeneracy and $g=g_sg_v=4$) to six (spin degeneracy is removed and $g=6$, which is equivalent to $\nu_e=3$ with spin degeneracy). This is possible due to the presence of electron valleys in TMD bilayers at $\Lambda$ points. An important role is played by an increase in the thickness of the monolayer/bilayer heterostructure compared to the monolayer/monolayer one: the exciton has a lower binding energy than the EHL, because there is a gain in the ground state energy of EHL with a decrease in kinetic energy and an increase in the modulus of the correlation energy (the electrostatic, dipole-dipole and modulo exchange contributions also decrease). This paves the way for the experimental study of the charge-separated EHL in TMD-based heterostructures.

We have studied the gas--liquid phase transition in the system of electrons and holes in considered heterostructures. The phase diagram has a standard form. The critical temperature of this transition exceeds room temperature.

We have analytically calculated the shape of the EHL line in the photoluminescence spectrum. At low temperatures, the EHL line width is equal to the Fermi energy with good accuracy. At high temperatures (near room temperature), it can exceed twice the value of the corresponding Fermi energy, ``flooding'' the exciton line, which, apparently, will look like a peak against the background of the tail of the EHL line towards the continuum spectrum edge.

~

\textbf{Acknowledgments}

The author is grateful to S. G. Tikhodeev for the helpful discussion and valuable advice regarding this publication. The work was supported by the Foundation for the Advancement of Theoretical Physics and Mathematics ``BASIS'' (the project no. 20-1-3-68-1).

\end{document}